\documentclass[a4paper]{article}
\usepackage{amsthm}

\newtheorem{theorem}{Theorem}

\newtheorem{rem}{Remark}

\newtheorem{defin}{Definition}

\newtheorem{lemma}{Lemma}
\newtheorem{problem}{Problem}

\usepackage{latexsym}
\usepackage{amssymb}
\usepackage[dvips]{graphicx}
\usepackage[caption=false]{subfig}
\usepackage[english]{babel}
\usepackage[latin1]{inputenc} % To use characters such as é without typing \'e
\usepackage{color}
\usepackage{amsmath}
\usepackage{amsfonts}
\usepackage{enumerate}

\begin{document}

\title{On some geometric aspects of the class of hv-convex swtching components}

\author{Paolo Dulio~\thanks{Dipartimento di Matematica ``F. Brioschi'', Politecnico di Milano, Piazza Leonardo da Vinci $32$, I-$20133$ Milano, Italy}\and Andrea Frosini~\thanks{Universit\`a
di Firenze, Dipartimento di Sistemi e Informatica, Viale Morgagni
$65$, $50134$ Firenze, Italy}}
\maketitle

\begin{abstract}

In the usual aim of discrete tomography, the reconstruction of an unknown discrete set is considered, by means of projection data collected along a set $U$ of discrete directions. Possible ambiguous reconstructions can arise if and only if switching components occur, namely, if and only if non-empty images exist having null projections along all the directions in $U$. In order to lower the number of allowed reconstructions, one tries to incorporate possible extra geometric constraints in the tomographic problem, such as the request for connectedness, or some reconstruction satisfying special convexity constraints. In particular, the class $\mathbb{P}$ of horizontally and vertically convex connected sets (briefly, $hv$-convex polyominoes) has been largely considered.

In this paper we introduce the class of $hv$-convex switching components, and prove some preliminary results on their geometric structure. The class includes all switching components arising when the tomographic problem is considered in $\mathbb{P}$, which highly motivates the investigation of such configurations, also in view of possible uniqueness results for $hv$-convex polyominoes.

It turns out that the considered class can be partitioned in two disjointed subclasses of closed patterns, called windows and curls, respectively, according as the pattern can be travelled by turning always clockwise (or always counterclockwise), or points with different turning directions exist. It follows that all windows have a unique representation, while curls consist of interlaced sequences of sub-patterns, called $Z$-paths, which leads to the problem of understanding the combinatorial structure of such sequences.

We provide explicit constructions of families of curls associated to some special sequences, and also give additional details on further allowed or forbidden configurations by means of a number of illustrative examples.

\textbf{Keywords}: curl, Discrete Tomography, $hv$-convex set, polyomino, projection, switching-component, window, X-ray.\smallskip

\textbf{Math subject classification}: 52A30, 68R01, 52C30, 52C45

\end{abstract}

\section{Introduction}

Discrete Tomography is a part of the wider area of Computerized Tomography, which relates to a huge number of applications where image reconstruction from X-ray collected data is required. While Computerized Tomography involves analytical techniques and continuous mathematics (see, for instance \cite{KS,natterer}), Discrete Tomography is mainly concerned with discrete and combinatorial structures, it works with a small number of density values, in particular with homogeneous objects, and usually allows very few X-rays directions to be considered (see \cite{HK,HK2} for a general introduction to the main problems of Discrete Tomography).

The reconstruction problem is usually {\em ill-posed}, meaning that \textit{ambiguous reconstructions} are expected. To limit the number of allowed configurations, further information is usually incorporated in the tomographic problem, which sometimes leads to a unique solution (see for instance  \cite{GG}) in the case of convex reconstructions), or to the enumeration of the allowed solutions (an example with two projections is \cite{CFRR}).

In case different discrete sets $Y_1$ and $Y_2$ are \textit{tomographically equivalent} with respect to a set $U$ of directions, namely $Y_1, Y_2$ can be reconstructed by means of the same X-rays with respect to $U$, then there exist specific patterns, called \textit{switching components} which turn $Y_1$ into $Y_2$. Understanding the combinatorial and the geometric structure of the switching components is a main issue in discrete tomography (see, for instance \cite{AT,BDNP,BDFR,BDP1,DPa,FS,FV,GG,GE,HT,ryser2}).

A largely investigated case concerns the class $\mathbb{P}$ of $hv$-convex polyominoes, i.e., finite connected subsets of $\mathbb{Z}^2$ that are horizontally and vertically convex. Early results for two projections can be found in \cite{BDNP}, where a uniqueness conjecture has been also stated, later disproved in \cite{DFPR}. On this regard, a main role is played by switching components with respect to the horizontal and to the vertical directions, respectively denoted by $\overrightarrow{h}$ and $\overrightarrow{v}$. In \cite{GE}, such switching components have been studied from an enumerative and an algorithmic point of view, which provided a very interesting and illustrative presentation of their connection with the complexity of the reconstruction problem.  In this paper we also focus on such switching components, but we follow a different approach, based on a special geometrical condition (see Definition \ref{def:hv-convex switching component}), which defines a class of patterns called \textit{$hv$-convex switching components}. It includes the classes of \textit{regular} and of \textit{irregular} switching components considered in \cite{GE}, that we redefine in terms of $hv$-convex \textit{windows} and $hv$-convex \textit{curls}, respectively.

The geometric condition in Definition \ref{def:hv-convex switching component} is always satisfied when the switching component is determined by a pair of sets $Y_1, Y_2$ both internal to the class $\mathbb{P}$. This motivates a deep investigation of the structure of $hv$-convex switching components, in view of possible uniqueness results for $hv$-convex polyominoes.

We give a geometric characterization of $hv$-convex windows (Theorem\ref{teo:hvwindows}), and a necessary condition for a curl to be a $hv$-convex switching component (Theorem \ref{teo:nothvcurl}). In general, the condition is not sufficient, but it provides a basic information concerning the  geometric structure of $hv$-convex curls, which leads to the problem of understanding their geometric and combinatorial structure.

%We provide explicit constructions of special families of curls, and also give additional details on further allowed or forbidden configurations.

\section{Notations and preliminaries}

We first introduce some notations and basic definitions. As usual,  $\mathbb{R}^2$  denotes the Euclidean two-dimensional space, and $\mathbb{Z}^2\subset\mathbb{R}^2$ is the lattice of points having integer coordinates. If $A$ is a subset of $\mathbb{R}^2$, we denote by $int(A)$ and by $conv(A)$ the {\it interior} and the \textit{convex hull} of $A$, respectively. If $A$ consists of two distinct points $v$ and $w$, then $conv( A)$ is a \textit{segment}, denoted by $s(v,w)$.
%For any subspace $U\subset\mathbb{R}^2$, $U^{\perp}$ is its orthogonal complement.
If $A$ is a finite set of $\mathbb{Z}^2$, then $A$ is said to be a \textit{lattice set}, and $|A|$ denotes the number of elements of $A$. A {\em convex lattice set} is a lattice set $A\subset \mathbb{Z}^2$ such that $A=(conv(A))\cap\mathbb{Z}^2$.

%
% qui inserirei L_v(x) e L_h(x) per denotare la linea discreta orizz e vert per x
%

By $\overrightarrow{h},\overrightarrow{v}$ we mean the horizontal and the vertical directions, respectively. For any point $v\in \mathbb{R}^{2}$, we indicate by $L_h(v)$ and $L_v(v)$ the horizontal and the vertical line passing through $v$, respectively.

Finally, we define \emph{horizontal} (resp. \emph{vertical}) \emph{projection} of a finite set $A\subset\mathbb{Z}^2$ to be the integer vector $H(A)$ (resp. $V(A)$) counting the number of points of $A$ that lie on each horizontal (resp. vertical) line passing through it. We underline that such a notion of projection can be defined for a generic set of discrete lines parallel to a given (discrete) direction.

In literature, the word \textit{polyomino} indicates a connected finite discrete set of points. In particular, a polyomino is \textit{hv-convex} if each one of its rows and columns is connected. As it is commonly assumed, a polyomino is composed by rows and columns due to the habit of representing it by a binary matrix whose dimensions are those of its minimal bounding rectangle. The class of all $hv$-convex polyominoes is denoted by $\mathbb{P}$.

Given a point $v=(i,j)\in\mathbb{Z}^2$, the four following closed regions are defined (with the same notations as in \cite{qconvex,inscribable}):
\begin{eqnarray*}
Z_0(v)  = \{(i',j')\in\mathbb{R}^2 : i'\leq i,\, j'\leq j\},\:\:\:\:Z_1(v)  =  \{(i',j')\in\mathbb{R}^2 : i'\geq i,\, j'\leq j\},\\
Z_2(v) = \{(i',j')\in\mathbb{R}^2 : i'\geq i,\, j'\geq j\},\:\:\:\:Z_3(v)  = \{(i',j')\in\mathbb{R}^2 : i'\leq i,\, j'\geq j\}.
\end{eqnarray*}

A set of points $A$ is said to be \emph{Q-convex} (quadrant convex) along the horizontal and vertical directions if $Z_l(v)\cap A\neq\emptyset$
for all $l=0,1,2,3$ implies $v\in A$.

\begin{lemma}\label{lem:prisoner}
Let $P$ be a hv-convex polyomino, and consider a point $v\in\mathbb{Z}^2$. If $w_1,w_2,w_3\in P$ exist such that $Z_i(v)\cap\{w_1,w_2,w_3\}\neq\emptyset$ for all $i=0,1,2,3$, then $v\in P$.
\end{lemma}

\proof By \cite[Proposition 2.3]{qconvex}, a $hv$-convex set is also $Q$-convex with respect to the horizontal and to the vertical directions. The statement follows immediately by the hv-convex property of $P$.\qed\smallskip

\subsection{Switching components and the uniqueness problem}

\begin{defin}\label{def:switching component}
A pair $S=(S^0,S^1)$ of sets of points is a \textit{hv-switching} if:
\begin{itemize}
\item[-] $S^0 \cap S^1 = \emptyset$ and $|S^0| = |S^1|$;
\item[-] $H(S^0)=H(S^1)$ and $V(S^0)=V(S^1)$, i.e., $S^0$ and $S^1$ have the same horizontal and vertical projections.
\end{itemize}
\end{defin}

Each set $S^0$ and $S^1$ is indicated as $hv$-switching component.
We underline that also the notion of switching can be extended to the projections along a generic set of discrete directions (again refer to \cite{HK,HK2} for these definitions and the related main results).

A discrete set $A$ contains a $hv$-switching component if $S^0\subseteq A$ and $S^1\cap A=\emptyset$. In this case, we consider $A=Y\cup S^0$, with $Y$ being a (possibly void) discrete set; we define the set  $A'=Y\cup S^1$ as the \textit{dual} of $A$, and we say that the switching $S$ is \emph{associated} to $A$ and $A'$.

\subsection{$hv$-convex switching}\label{sub:hv-convex switching components}

A classical result in \cite{ryser2} states that if $A_1$ and $A_2$ are two discrete sets sharing the same horizontal and vertical projections, then $A_2$ is the dual of $A_1$ with respect to a $hv$-switching. So, for any point $v\in S^0$ (resp. $v\in S^1$), there exist points $w_1,w_2\in S^1$ (resp. $w_1,w_2\in S^0$) such that $w_1\in L_h(v)$ and $w_2\in L_v(v)$.

If the sets $A_1$ and $A_2$ are $hv$-convex polyominoes, then, due to Lemma \ref{lem:prisoner}, for any $x\in S$ there exists one and only one $i\in\{0,1,2,3\}$ such that $Z_i(x)\cap S$ consists of points all belonging to the same component of $S$ as $x$. The quadrant $Z_i(x)$ is said to be the \textit{free region} of $x$, or the \textit{$S$-free region} of $x$ in case we wish to emphasize that the free region relates to the switching $S$. We denote by $F(x)$ (or by $F_S(x)$) the free region of $x\in S$. Also, $F_i(S)$ denotes the subset of $S$ consisting of all points having free region $Z_i(x)$, namely $F_i(S)=\{x\in S,\:F_S(x)=Z_i(x)\}$, $i\in\{0,1,2,3\}$.\smallskip

We have the following

\begin{lemma}\label{lem:cond2}
Let $S=(S^0,S^1)$ be a $hv$-switching. Then, the following conditions are equivalent

\begin{equation}\label{eq:hv-1}
\displaystyle{\bigcup_{i=0}^3Z_i(S)=S.}
\end{equation}

\begin{equation}\label{eq:hv-2}
v,w\in Z_i(S), i\in\{0,1,2,3\}, v\in S^0, w\in S^1 \Rightarrow\:v\notin Z_j(w),\:w\notin Z_j(v), j=i+2\:(mod\: 4).
\end{equation}

\end{lemma}

\proof  Let $v,w\in S$ such that $v,w\in Z_i(S)$,  with $v\in S^0, w\in S^1$. Suppose that $v\in Z_j(w)$, with $j=i+2\:(mod\: 4)$. Then $w\in Z_i(v)$, a contradiction. Analogously, if $w\in Z_j(v)$, with $j=i+2\:(mod\: 4)$, then $v\in Z_i(w)$, a contradiction. Therefore, \eqref{eq:hv-2} holds. Conversely, assume that \eqref{eq:hv-2} holds. Let $v\in S$, and suppose $v\in S^0$. Since $S$ is a $hv$-switching, then there exist three values of $k\in\{0,1,2,3\}$ such that $Z_k(v)\cap S^1\neq\emptyset$. Suppose that $w\in S^1$ exists such that $w\in Z_i(v)$ for $i\neq k$. Then $v\in Z_j(w)$, where $j=i+2\:(mod\:4)$, which contradicts \eqref{eq:hv-2}. Therefore, $v$ has a free region, namely $F(v)=Z_i(v)$. With the same argument we get that any $w\in S^1$ has a free region. Therefore, each point of $S$ has a nonempty free region, and \eqref{eq:hv-1} follows.\qed\smallskip

\begin{defin}\label{def:hv-convex switching component}
Let $S=(S^0,S^1)$ be a $hv$-switching. Then, $S$ is said to be a \textit{$hv$-convex switching} if one of the equivalent conditions of Lemma \ref{lem:cond2} holds.
\end{defin}

\begin{rem}\label{rem:hv-convex condition}
By the above discussion, if $S=(S^0,S^1)$ is a $hv$-switching associated to a pair of $hv$-convex polyominoes, then \eqref{eq:hv-1} holds, so $S$ is a $hv$-convex switching. However, the converse is not necessarily true, namely it could exist two polyominoes $P_1$ and $P_2$ that are one the dual of the other with respect to $S$ and such that one or both of them are not $hv$-convex polyominoes. An interesting case is Figure $23$ in \cite{GE}, or Figure \ref{fig:hv_window_Regions} below.
\end{rem}

%In the sequel, we focus our interest on the problem of characterizing the class of $hv$-convex switchings.% Variants of this problem attracted the researchers' interest since the beginning of Discrete Tomography, never loosing the appeal in the community, as recently witnessed in \cite{GE} and similar works.

\subsection{Squared spirals} A closed polygonal curve $K$ in $\mathbb{R}^2$ is said to be a \textit{squared spiral} if $K$ consists of segments having, alternatively, horizontal and vertical direction. Their endpoints form the \textit{set of vertices} of the polygonal, denoted by $V(K)$. Two squared spirals are said to \textit{intersect} in case some of (possibly all) their segments done. Assume to travel $K$ according to a prescribed orientation. A vertex $v$ of $K$ is said to be a \textit{counterclockwise point} if, crossing $v$, implies a counterclockwise change of direction. Differently, $v$ is a \textit{clockwise point}. Of course, by reversing the travelling orientation, clockwise and counterclockwise vertices mutually exchange. The \textit{bounding rectangle} of $K$ is the smallest rectangle $R_K$  containing $K$.

\subsection{Windows and curls} We now introduce two classes of special squared spirals that provide a geometric reformulation of the notions of \textit{regularity} and of \textit{irregularity} discussed in \cite{GE}, which, in addition, constitute the main focus of our study. A squared spiral $W$ is said to be a \textit{window} if it can be traveled by turning always clockwise, or always counterclockwise. Differently, the squared spiral is said to be a \textit{curl}. Therefore, travelling a curl needs changes of turning direction.

%A window is said to be a \textit{square} if it has no self intersections, i.e., it consists of four sides only.

%In the sequel, we will consider windows and curls whose vertices lie in the integer lattice only.

%
%\begin{figure}[htbp]
%\centering
%\includegraphics[scale=0.20,viewport=50 550 550 750,clip=true]{curl_window.jpg}
%\caption{Examples of squared spirals. A window on the left and a curl on the right.}
%\label{fig:curl_window}
%\end{figure}

Obviously a rectangle is a particular case of window that coincides with its bounding rectangle.

\begin{rem}\label{rem:switching} Each window and each curl form a $hv$-switching $S=(S^0,S^1)$ by considering the corresponding vertices alternatively belonging to $S^0$ and $S^1$.
\end{rem}

%{\color{red} qui metterei il lemma: i vertici di una window o di un curl, considerati alternativamente, formano una hv switching component.}

\subsection{$Z$-paths}\label{sub:Z_paths} 
\begin{figure}[htbp]
\centering
\includegraphics[scale=0.40,viewport=50 640 550 715,clip=true]{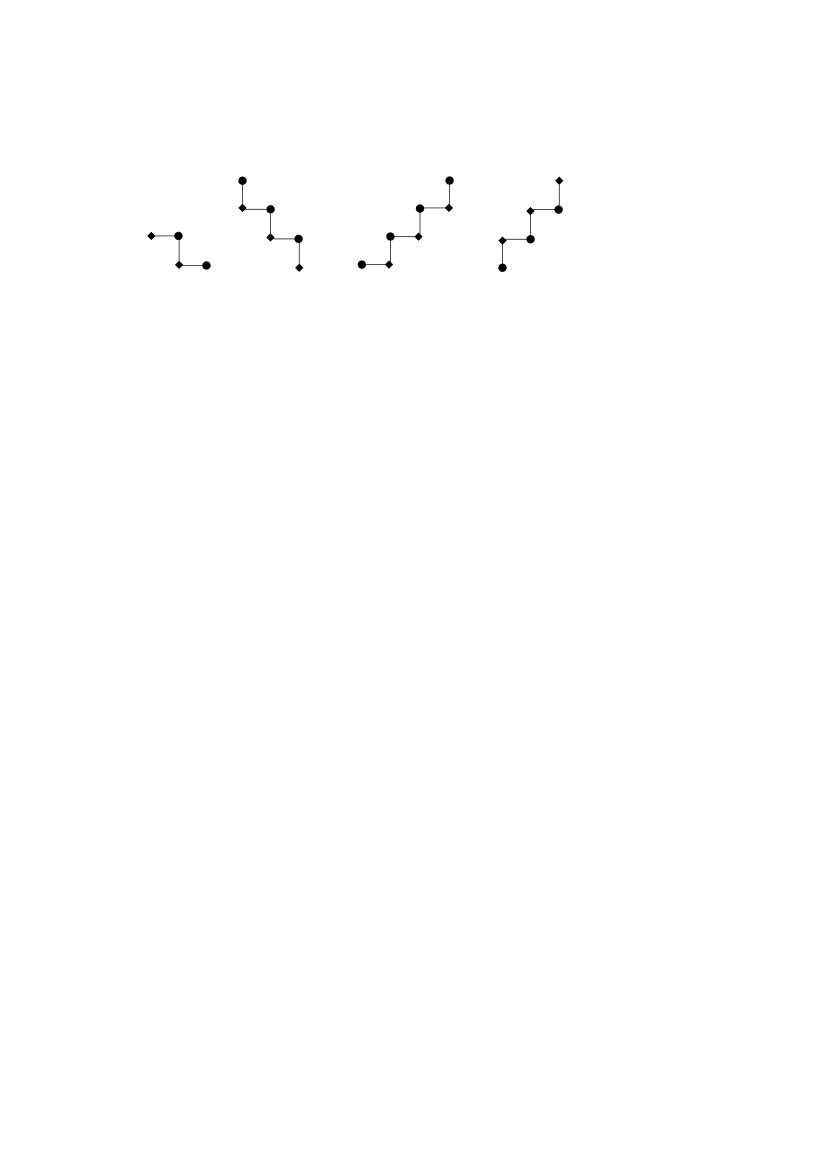}
\caption{Different types of $Z$-paths. From left to right: A simple $hvh$ SE-NW $Z$-path, a $v(hv)_2$ SE-NW $Z$-path, a $(hv)_3$ SW-NE $Z$-path, and a $v(hv)_2$ SW-NE $Z$-path.}
\label{fig:typeZ_paths}
\end{figure}

A $Z$-path is a staircase shaped pattern consisting of a monotone sequence of horizontal and vertical segments, whose vertices alternate between clockwise and counterclockwise points. We say that the $Z$-path is of \textit{type} SE-NW, or SW-NE, according as it can be travelled moving from South-East to North-West (or conversely), or from South-West to North-East (or conversely), respectively. A \textit{simple}, or \textit{one-level}, $Z$-path consists of just three segments, horizontal-vertical-horizontal, or vertical-horizontal-vertical, referred to as $hvh$, or $vhv$ $Z$-path, respectively. Excluding its endpoints, a simple $Z$-path exhibits a pair of vertices having a specified orientation, clockwise-counterclockwise, or clockwise-counterclockwise, according to the considered type, and moving from south to north along the pattern. In general, for $q>0$, we have a \textit{$q$-level} $Z$-path if, excluding its endpoints, it consists of $q+1$ vertices having alternating orientations. Therefore, if $q$ is odd, we have $q$ horizontal and $q-1$ vertical segments, or conversely, and we refer to the corresponding $Z$-path with the notation $h(vh)_{q-1}$ and $v(hv)_{q-1}$, respectively. If $q$ is even, then the $Z$-path consists of $q$ horizontal and of $q$ vertical segments, and we adopt the notation $(hv)_q$, or $(vh)_q$, according as the first segment is horizontal or vertical (see Figure \ref{fig:typeZ_paths}). Any $Z$-path is a $hv$-convex set. In a SE-NW $Z$-path, any vertex $v$, different from an endpoint, has free region $Z_0(v)$ or $Z_2(v)$, while, in a SW-NE $Z$-path, the free region is $Z_1(v)$ or $Z_3(v)$. In any case, the elements of the sets of free regions $\{Z_0(v), Z_2(v)\}$, or $\{Z_1(v), Z_3(v)\}$ alternate along the $Z$-path. Since the vertices of a $Z$-path are, alternatively, clockwise and counterclockwise oriented, then no $q$-level $Z$-path, with $q>0$, can be found in a window, while any curl surely includes some $Z$-paths. Differently, if $q=0$, we have an $L$-shaped path, consisting of an horizontal and a vertical segment, with just one intermediate point. We refer to such a path as a \textit{degenerate $Z$-path}. Note that a window can be considered as a consecutive sequence of degenerate $Z$-paths, while, in a curl, different $Z$-paths (possibly degenerate) can appear. In what follows, we provide a precise characterization of how these paths can be combined together.

\section{Characterization of $hv$-convex windows}\label{sec:Characterization of $hv$-convex windows and $hv$-convex curls}

We give a necessary and sufficient condition for a window to be $hv$-convex. This leads to a geometric characterization of the $hv$-convex switchings that have the structure of a  $hv$-convex window.\smallskip

\begin{theorem}\label{teo:hvwindows}
Let $W$ be a window of size $n\geq 1$ and $\{w_1,w_2,...,w_{4n}\}$ be the set of its vertices. Then $W$ is a $hv$-convex switching if and only if a point $x\in\mathbb{R}^2$ exists such that $w_i\in Z_0(x)\cup Z_2(x)$ for all the odd indices, and $w_i\in Z_1(x)\cup Z_3(x)$ for all the even indices.
\end{theorem}

\proof Assume that a point $x\in\mathbb{R}^2$ exists such that $w_i\in Z_0(x)\cup Z_2(x)$ for all the odd indices, and $w_i\in Z_1(x)\cup Z_3(x)$ for all the even indices. Then, by definition of window, $W$ has the same number of vertices in each $Z_i(x)$, $i=0,1,2,3$, namely, $w_i\in Z_0(x)$, for $i=1\:(mod\:4)$, $w_i\in Z_1(x)$, for $i=2\:(mod\:4)$, $w_i\in Z_2(x)$, for $i=3\:(mod\:4)$, and $w_i\in Z_3(x)$, for $i=0\:(mod\:4)$. Therefore, if $W^0$ and $W^1$ are, respectively, the set of the even and of the odd labeled vertices of $W$, then each point of $W^0$ has a horizontal and a vertical corresponding in $W^1$ and conversely. This implies that the free regions of all points in $W^0$ are contained in $Z_1(x)$ or in $Z_3(x)$, and the free regions of all points in $W^1$ are contained in $Z_0(x)$ or in $Z_2(x)$. Therefore, $W$ is $hv$-convex.

%{\color{red} se inseriemo il lemma che dice che ogni windows e ogni curl e' una hv switching allora la prova si semplifica. Riguardare la dimostrazione!}

Conversely, suppose that $W$ is a $hv$-convex switching. Without loss of generality we can assume that $W$ is traveled counterclockwise, starting from $w_1$. Also, up to a rotation (which does not change the argument) we can always assume that the free region of $w_1$ is $Z_0(w_1)$. Then the free region of $w_i$ is $Z_j(w_i)$ where $i-j=1\:(mod\:4)$. For $j=0,1,2,3$, let $H_j$ be the set

$$H_j=\bigcup_{i=j+1\:(mod\:4)}Z_j(w_i).$$

\begin{figure}[htbp]
\centering
\includegraphics[scale=0.40,viewport=100 520 350 740,clip=true]{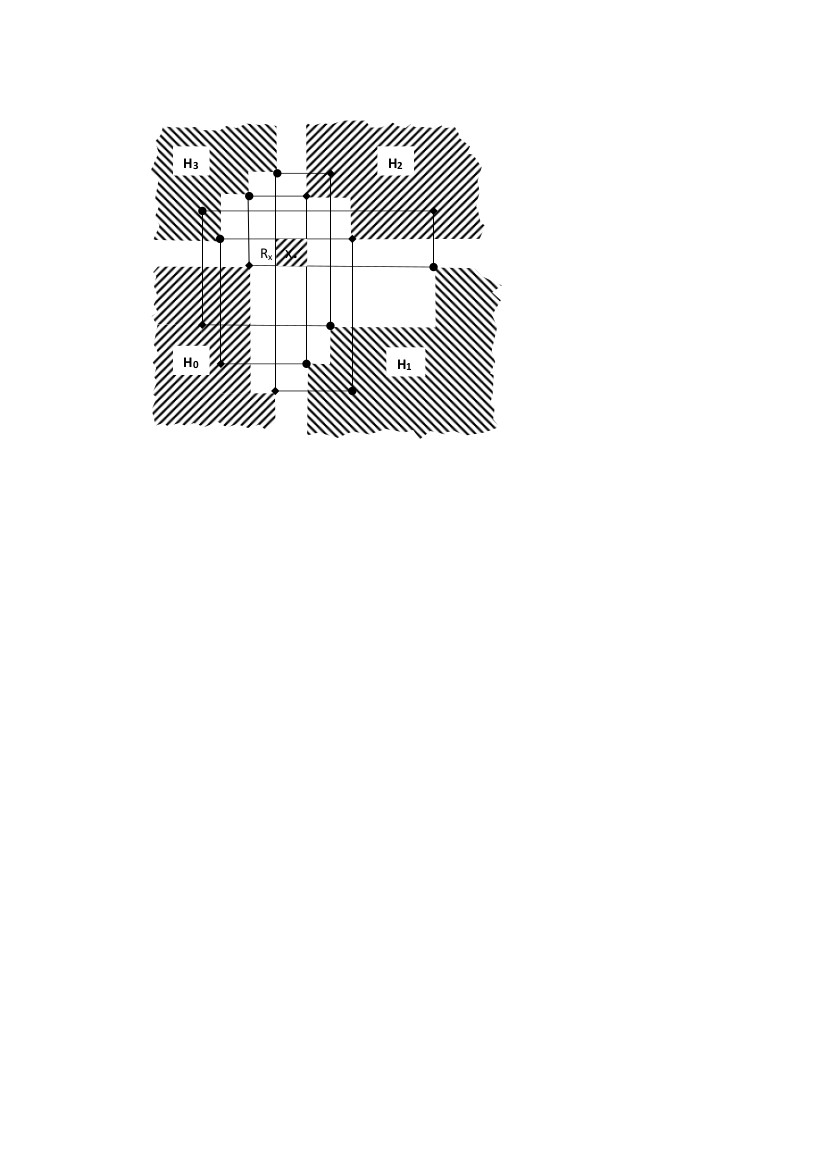}
\caption{The four regions $H_i$, $i\in\{0,1,2,3\}$ related to a $hv$-convex window. The rectangle $R_x$ contains all the points having the property stated in Theorem~\ref{teo:hvwindows}.}
\label{fig:hv_window_Regions}
\end{figure}

Due to the $hv$-convexity of $W$, the sets $H_i$ are mutually disjointed. Consider the strip bounded by the two horizontal lines supporting $H_0\cup H_1$ and $H_2\cup H_3$, and the strip bounded by the two  vertical lines supporting $H_0\cup H_3$ and $H_1\cup H_2$ (see Figure \ref{fig:hv_window_Regions}).

The intersection of such strips forms a rectangle $R$, having horizontal and vertical sides, and with no points of $W$ belonging to the internal $int(R)$ of $R$. Let $x$ be any point such that $x\in int(R)$. Then $H_j\subseteq Z_j(x)$ for all $j=0,1,2,3$, and the statement follows.\qed\smallskip

\begin{rem}\label{rem:size} The property stated in Theorem~\ref{teo:hvwindows} is not restricted to a single point, but it involves all the points belonging to the interior of the rectangle $R$.
\end{rem}\smallskip

For any $hv$-convex window $W$, and point $x\in\mathbb{R}^2$ as in Theorem \ref{teo:hvwindows}, all quadrants $Z_i(x)$, $i\in\{0,1,2,3\}$ contain the same number of points of $W$, which is said the \textit{size} of the window. Note that a window can be a switching component with respect to the horizontal and vertical directions without being $hv$-convex.

%, as in Figure~\ref{fig:curl_window}, where no $x$ as in Theorem~\ref{teo:hvwindows} exists.

\section{Characterization of $hv$-convex curls}

Moving to curls, a deeper analysis is required, as it has been pointed out in~\cite{GE} in terms of irregular switching components. Here we push the study a step ahead, by investigating the geometric nature and the main features of those curls that form $hv$-convex switching. As a first result, we prove a necessary condition for a curl to be a $hv$-convex switching, say \emph{hv-convex curl}. In general, the given condition is not sufficient, but it spreads light on the geometric structure of the $hv$-convex curls, and leads to their characterization in terms of $Z$-paths. As a consequence, the class of $hv$-convex curls will be partitioned into two subclasses.\smallskip

\begin{theorem}\label{teo:nothvcurl}
Let $C$ be a curl that forms a $hv$-switching, and let $v$ and $w$ be two points in $V(C)$ with the same orientation. If precisely $2n>0$ consecutive vertices between $v$ and $w$ exist, and having their opposite orientation, then $C$ is not a $hv$-convex curl.
\end{theorem}

\proof Suppose that $C$ is $hv$-convex. Without loss of generality, we can assume that travelling $C$ from $v$ to $w$ the vertices $v$ and $w$ are counterclockwise oriented. Up to a rotation we can also assume that $Z_0(v)$ is the free region of $v$, so that a vertex $v_1\in Z_2(v)\cap Z_3(v)$ exists, with $v, v_1$ in different components of $C$. Let $x_1,...,x_{2n}$ be the $2n>0$ clockwise oriented vertices of $C$ that are crossed when moving from $v$ to $w$.

\begin{figure}[htbp]
\centering
\includegraphics[scale=0.40,viewport=50 580 400 750,clip=true]{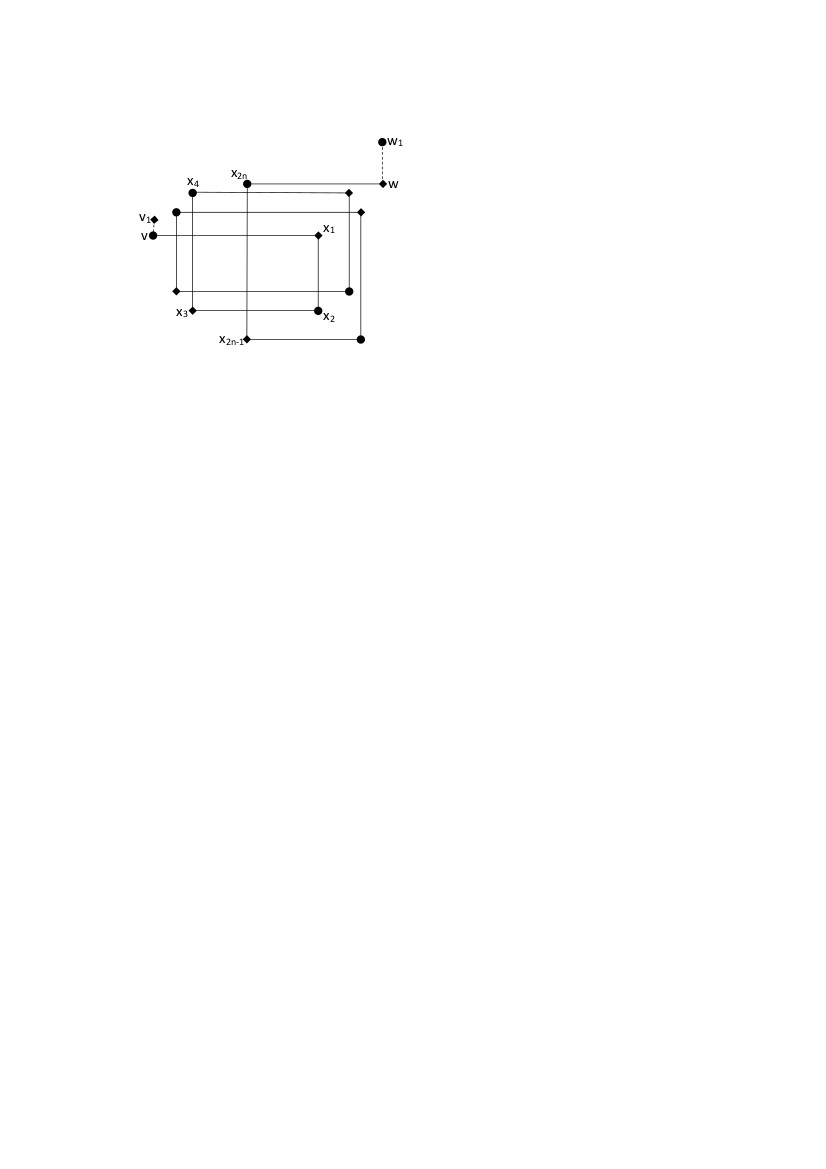}
\caption{Positions of consecutive vertices having a same orientation in a curl.}
\label{fig:non_hv_convex_curl}
\end{figure}

The segment $s(v, x_1)$ is horizontal. The same holds for the segment $s(x_{2n}, w)$, and also for all segments $s(x_{2k}, x_{2k+1})$, for $1\leq k\leq n-1$. Analogously, all segments $s(x_{2k-1}, x_{2k})$, for $1\leq k\leq n$  are vertical. Then $Z_2(x_1)$ is the free region of $x_1$, $Z_1(x_2)$ is the free region of $x_2$, $Z_0(x_3)$ is the free region of $x_3$, and, in general, the free region of $x_i$ is the quadrant $Z_j(x_i)$ such that $i+j=3\:(mod\: 4)$. Therefore, the free region of $x_{2n}$ is $Z_1(x_{2n})$ if $n$ is odd, and $Z_3(x_{2n})$ if $n$ is even, which implies that the free region of $w$ is, respectively, $Z_3(w)$ and $Z_1(w)$ (see Figure \ref{fig:non_hv_convex_curl}, where the case $F(w)=Z_1(w)$ is represented).

Now, all the vertices $x_h$, with $h$ odd, belong to a component different from that $v$, so they do not lie in $F(v)=Z_0(v)$. Since $x_1\in Z_1(v)\cap Z_2(v)$, then $x_2\in int(Z_1(v))$. If $n=1$, then $w\in int(Z_1(v))$, and $F(w)=Z_3(w)$, so that $v\in F(w)$, a contradiction, since $v$ and $w$  belong to different components of $C$. So, the statement follows for $n=1$. If $n>1$, then $x_3\in int(Z_1(v))$, so $x_4\in Z_1(v)\cup Z_2(v)$. However, $x_4\notin Z_1(v)$, since, otherwise $v_1\in F(x_4)=Z_3(x_4)$, a contradiction, being $v$ and $v_1$ in different components of $C$. By iterating the argument, we get that all the vertices of the form $x_{2k}$, with $k\leq n$ and $k$ even must belong to $Z_2(v)$. Analogously, all the vertices of the form $x_{2k}$, with $k\leq n$ and $k$ odd must belong to $Z_1(v)$, since, differently, $x_{2k}$ would belong to $F(x_1)=Z_3(x_1)$, or conversely, $x_1\in F(x_{2k})=Z_1(x_{2k})$, a contradiction, since $x_1$ and $x_{2k}$ belong to different components of $C$. This implies that the vertices $x_{2k-1}$ with $2\leq k\leq n$ and $k$ even belong to $Z_1(v)$, while the vertices $x_{2k-1}$ with $1<k\leq n$ and $k$ odd belong to $Z_2(v)$. Consequently, also $w\in Z_1(v)\cup Z_2(v)$.

Suppose that $w\in Z_1(v)$. Since the segment $s(x_{2n}, w)$ is horizontal, then $x_{2n}$ also belongs to $Z_1(v)$. As shown above, this implies that $n$ is odd, so $F(w)=Z_3(w)$, and consequently $v\in F(w)$, a contradiction.

Hence $w\in Z_2(v)$, then $x_{2n}$ also belongs to $Z_2(v)$, which implies that $n$ is even, and consequently $F(x_{2n})=Z_3(x_{2n})$, and $F(w)=Z_1(w)$. Therefore, $w$ must belong to $Z_2(x_2)$, otherwise $w\in Z_3(x_2)$, and consequently $x_2\in F(w)=Z_1(w)$, a contradiction. From $w\in Z_2(x_2)$, and $w\in Z_2(v)$, it follows that $w\in Z_2(x_1)$. Since $C$ is a switching with respect to the vertical direction, then there exists a vertex $w_1\in Z_2(w)\cap Z_3(w)$, with $w$ and $w_1$ in different components, and consequently also $x_1$ and $w_1$ belong to  different components of $C$ (see Figure \ref{fig:non_hv_convex_curl}). Since $w\in Z_2(x_1)$, then also $w_1\in F(x_1)=Z_2(x_1)$, a contradiction.

Consequently, the assumption that $C$ is $hv$-convex  always leads to a contradiction, and the statement follows.\qed\smallskip

\section{On some sequences associated to $hv$-convex switchings}

Let $S$ be a squared spiral. We associate to $S$ an integer sequence $(k_1, k_2,...,k_n)$, say \textit{$hv$-sequence}, where each $k_i$ represents the $i$-th maximal sequence of $k_i$ vertices that can be travelled clockwise or counterclockwise, with $i=1,2,...,n$. The starting vertex is not indicated, so the sequence can be considered up to circular shifts. If the sequence $(k_1, k_2,...,k_n)$ is periodic, then we adopt the notation $(k_1,...,k_{n'})_h$, to represent the $h$ time repetition of the sequence $(k_1,...,k_{n'})$, with $n=n' \cdot h$; if $h=1$, we choose to omit it. We are interested in characterizing the $hv$-sequences that admit a $hv$-convex switching, say {\em $hv$-convex sequences}. Therefore, we are led to the following general problem.

\begin{problem}\label{prob:sequences}
For which $k_1,...,k_n, h\in\mathbb{N}$ there exists a $hv$-convex sequence $(k_1,...,k_n)_h$?
\end{problem}\smallskip

Concerning windows, Problem \ref{prob:sequences} has an easy solution.

\begin{theorem}\label{teo:window_sequence}
For each $n>0$, $(4n)$ is a $hv$-convex sequence if and only if the associated square spiral is a window.
\end{theorem}

\proof By Theorem \ref{teo:nothvcurl}, the $hv$-sequence associated to a curl is of the form $(k_1,...,k_{n'})_h$, where $k_1,...,k_n$ are odd. Therefore, $(4n)$ cannot be the $hv$-sequence associated to a curl. Let $W$ be a window. By Theorem \ref{teo:hvwindows}, $x\in\mathbb{R}^2$ exists such that all quadrants $Z_i(x)$, $i\in\{0,1,2,3\}$ contain the same number of points of $W$.  Then, $W$ has $4n$ vertices, for some $n>0$, which implies that the $hv$-sequence associated to $W$ is $(4n)$.\qed\smallskip

Differently, Problem \ref{prob:sequences} seems to require a deeper investigation of the geometrical and combinatorial structure of the set of vertices of a curl. In this view, we give here some preliminary remarks. First of all, note that, having a $(k_1,...,k_n)_h$ curl, is in general not sufficient to get $hv$-convexity, since the conditions in Lemma~\ref{lem:cond2} do not automatically hold.

%Figure~\ref{fig:curl_3_7_non_convex} shows a $(3,7)_1$ curl that is not $hv$-convex is given.
%
%
%\begin{figure}[htbp]
%\centering
%\includegraphics[scale=0.4,viewport=50 550 250 710,clip=true]{curl_3_7_non_convex.jpg}
%\caption{A $(3,7)_1$ curl which is not hv-convex.}\label{fig:curl_3_7_non_convex}
%\end{figure}

\begin{figure}[htbp]
\centering
\subfloat[]
{\includegraphics[scale=0.5,viewport=100 650 200 720,clip=true]{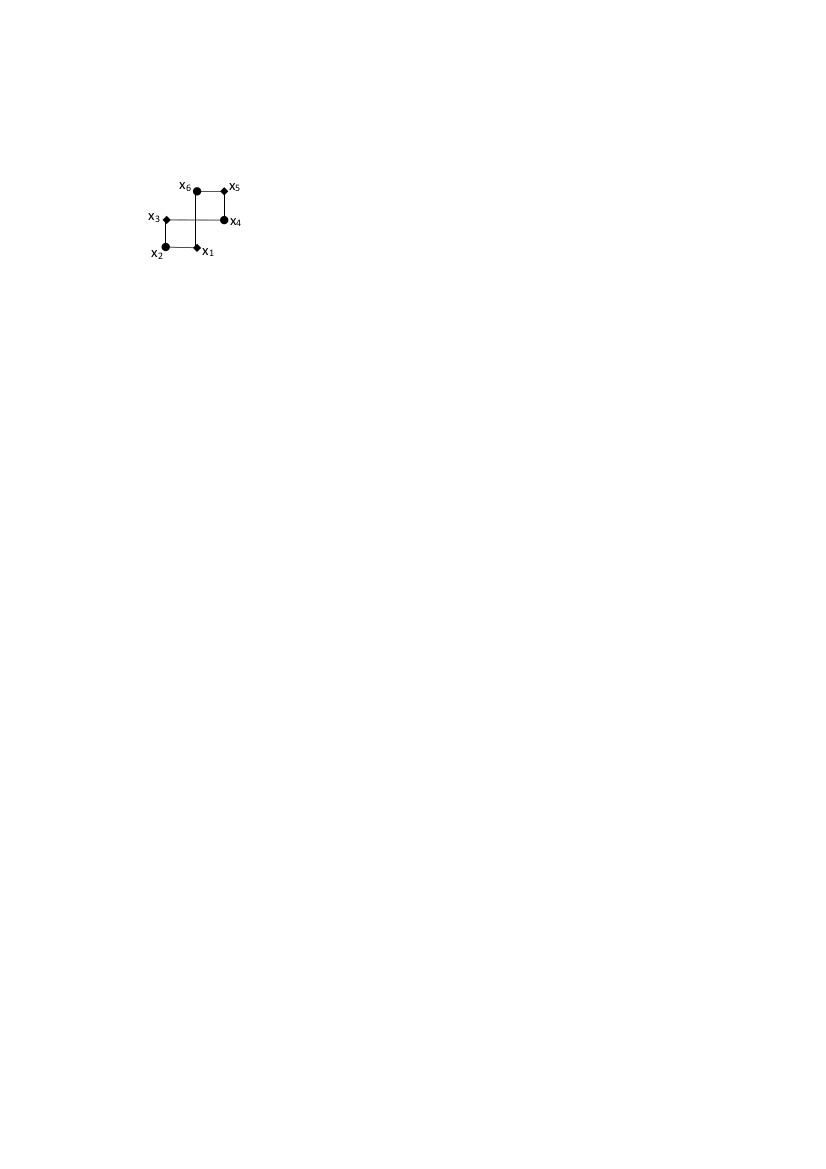}}
\centering
\subfloat[]
{\includegraphics[scale=0.5,viewport=50 540 220 685,clip=true]{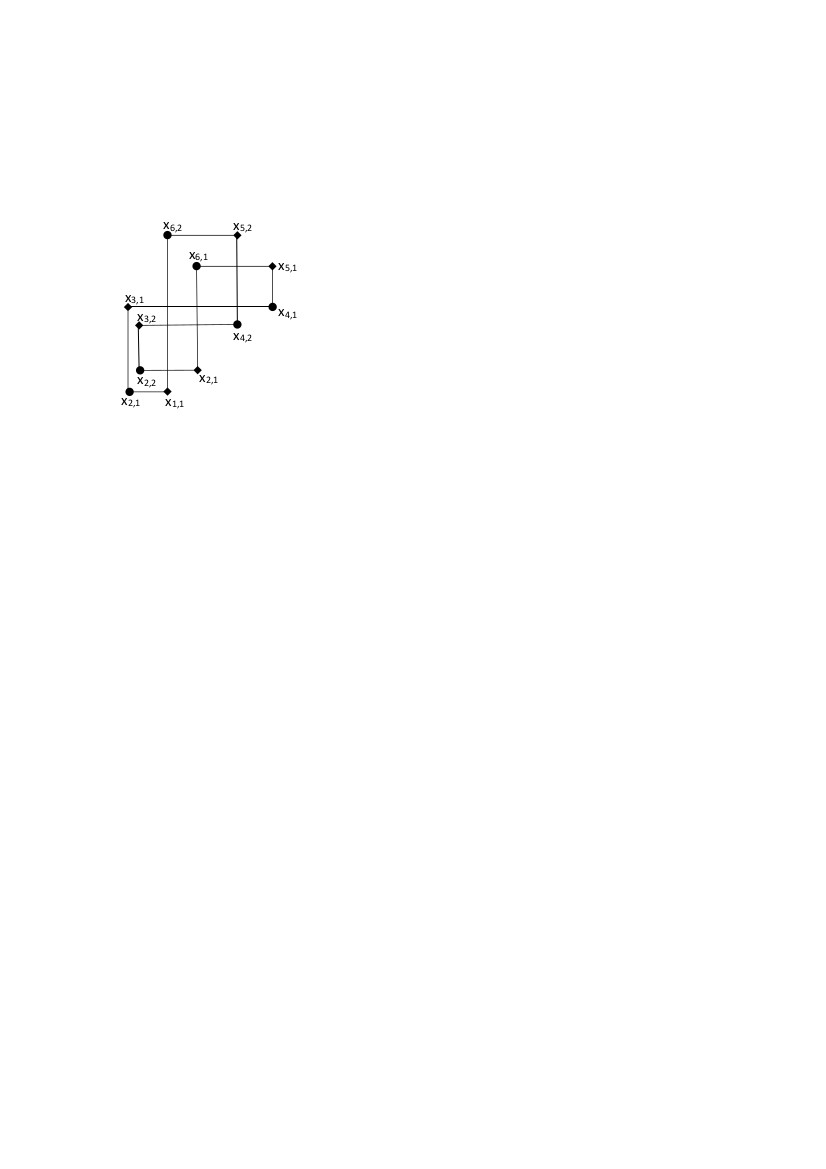}}
\centering
\subfloat[]
{\includegraphics[scale=0.40,viewport=55 560 350 750,clip=true]{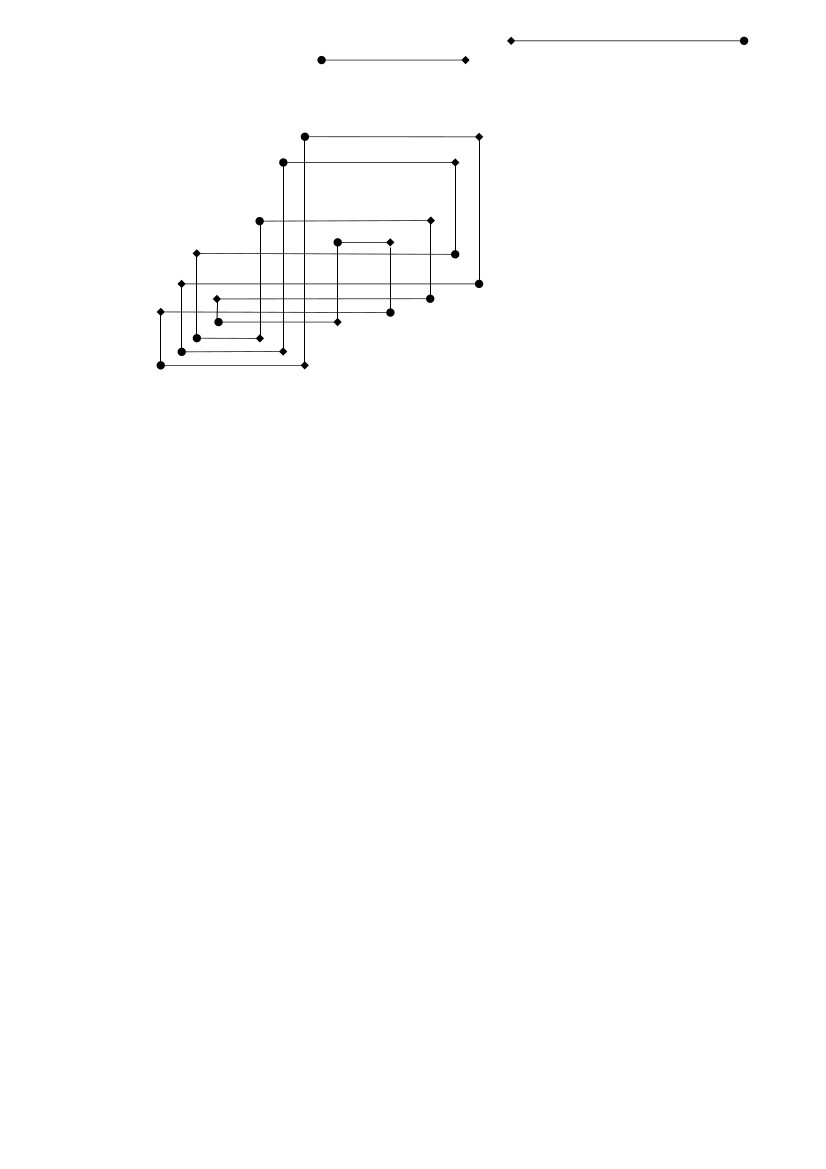}}
\caption{$(a)$ A $(3,3)_1$ $hv$-convex curl $C$, corresponding to a SW-NE vertex-gluing of two rectangles. $(b)$ A $(3,3)_2$ $hv$-convex curl with two pairs of intersecting simple $Z$-paths. $(c)$ Example of curl associated to the $hv$-sequence $(3,3)_4$.}
\label{fig:simple_curls}
\end{figure}

 A vertex $v$ of a curl whose turning direction differs from that of the previous encountered point, is said to be a \textit{changing point}. In order to improve our knowledge on the $hv$-sequences allowed for curls, it is worth focusing on the possible $Z$-paths that can be included in a $hv$-convex curl, which reflects in the understanding of how changing points can occur. As already observed, the simplest $hv$-convex curl is the $(3,3)_1$ curl shown in Figure~\ref{fig:simple_curls} $(a)$. Its vertices consists of six points $x_1, \dots, x_6$, where $F(x_1)=Z_1(x_1)$, $F(x_2)=Z_0(x_2)$, $F(x_3)=Z_3(x_3)$, $F(x_4)=Z_1(x_4)$, $F(x_5)=Z_2(x_5)$, $F(x_6)=Z_3(x_6)$. The $(3,3)_1$ curl can also be considered as the join of two simple $hvh$ and $vhv$ SW-NE $Z$-paths (see Section \ref{sub:Z_paths}) having $x_2$ and $x_5$ in common. This means that $x_1, x_4$ are changing points (or $x_3$, $x_6$, depending on the starting choice for the walking direction). Figure~\ref{fig:simple_curls} $(b)$ shows a $(3,3)_2$ $hv$-convex curl, consisting of two different pairs of intersecting simple SW-NE $Z$-paths (analogous constructions can be performed by using SE-NW $Z$-paths).
Analogously, for any integer number $h\geq 1$, a curl $C$ can be constructed having $h$ pairs of intersecting SW-NE $Z$-paths. These can be consecutively arranged, or, differently, connected by means of L-shaped paths, as described above. See Figure~\ref{fig:simple_curls} $(c)$ for an example where $h=4$.

%{\color{red} potremmo mettere anche in questa figura (a) (b) e (c) mantenendo la notazione delle altre. Non potremmo inoltre eliminare una delle figure 6 e 7?}

%
%\begin{figure}[htbp]
%\centering
%\includegraphics[scale=0.40,viewport=55 560 500 750,clip=true]{simple_curl_4.jpg}
%\caption{Example of curl associated to the $hv$-sequence $(3,3)_4$.}
%\label{fig:simple_curl_4}
%\end{figure}

Of course, curl containing $Z$-paths of higher level can be also constructed.

\begin{figure}[htbp]
\centering
\includegraphics[scale=0.40,viewport=50 610 550 750,clip=true]{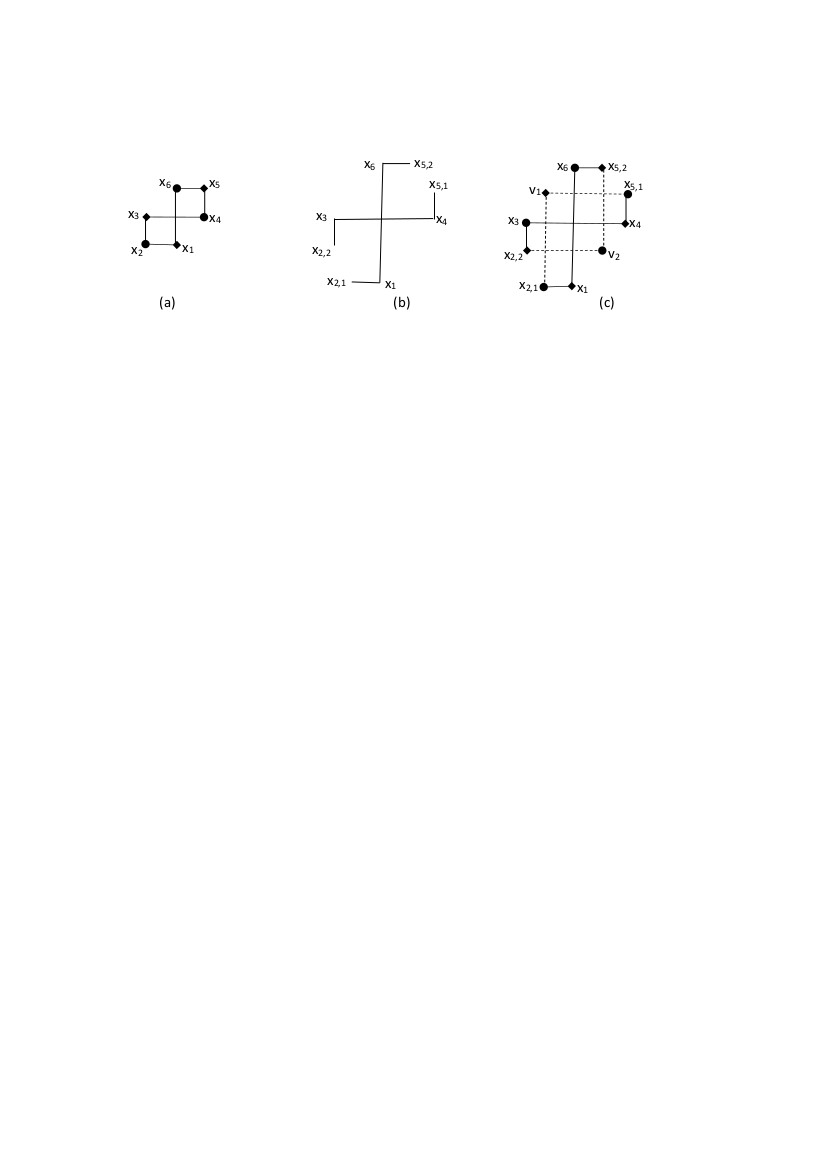}
\caption{Including L-shaped paths in a given curl. (a) The starting curl. (b) The split of the two constituent simple $Z$-paths. (c) The connection of the two simple $Z$-paths by joining their extremal vertices with two degenerate $Z$-pats.}
\label{fig:curltransformation}
\end{figure}

However, different $Z$-paths of a same curl are not necessarily consecutive. For instance, Figure~\ref{fig:curltransformation} shows how to insert degenerate $Z$-paths ($L$-shaped paths) between the bottom-left endpoint of a $Z$-path and the upper-right endpoint of a different $Z$-path, so transforming a $(3,3)_1$ curl into a $(5,5)_1$ curl.

\begin{figure}[htbp]
\centering
\includegraphics[scale=0.40,viewport=100 600 300 785,clip=true]{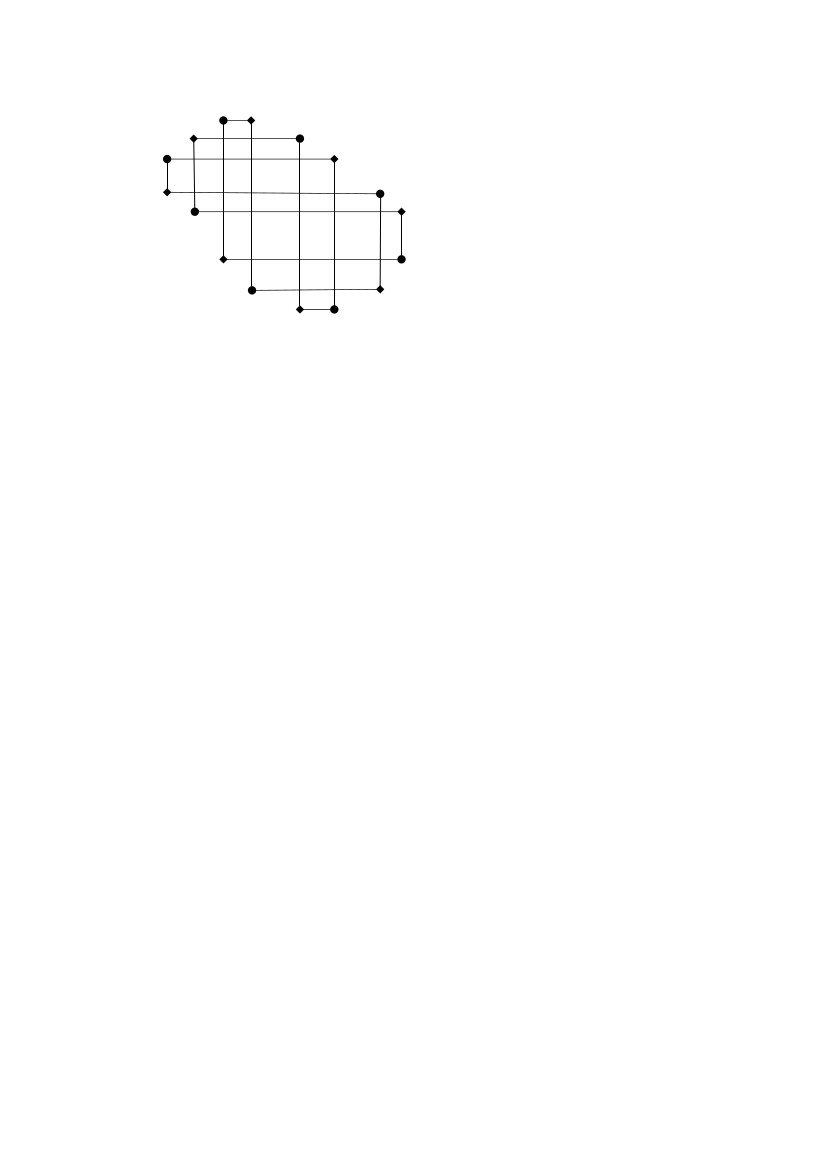}
\caption{The hv-convex curl associated to the $hv$-convex sequence $(3,5)_2$.}\label{fig:curl_3_5_convex}
\end{figure}

Further constructions also exist having associated $hv$-convex sequence of type $(k_1,k_2)_h$, with $k_1\neq k_2$. Figure~\ref{fig:curl_3_5_convex} shows a curl associated to the $hv$-convex sequence $(3,5)_2$.

\section{Conclusion and remarks}
We have introduced the class of $hv$-convex switching components, which includes all switching components associated to a pair of tomographically equivalent $hv$-convex polyominoes. We have separated the class in two disjointed subclasses of closed patterns, the windows and the curls, respectively. We have given geometrical results on both subclasses, which leads to the problem of characterizing them in terms of $hv$-convex sequences. While windows provide a complete and easy solution, deeper investigation is required for curls. We have discussed a few preliminary allowed or forbidden $hv$-sequences, which  provide partial answers to Problem \ref{prob:sequences} in the case of curls. For a complete solution to Problem \ref{prob:sequences} it becomes relevant to understand how, in general, different $Z$-paths can be connected between them in a same curl. In particular, it would be worth exploring possible connections between the allowed levels of the $Z$-paths in a same curl, and the degree of convexity of $L$-convex sets \cite{CFRR}. We wish to investigate in these directions in separated further works

\end{document}